\documentclass[aps,prb,twocolumn,groupedaddress]{revtex4-1}
\usepackage{color,bm,amsthm,amssymb,mathtools}

\begin{document}

\title{Persistence of pinning and creep beyond critical drive within the strong pinning paradigm}

\author{M.\ Buchacek$^1$}
\author{R.\ Willa$^{1,2}$}
\author{V.B.\ Geshkenbein$^1$}
\author{G.\ Blatter$^1$}
\affiliation{$^1$Institute for Theoretical Physics, ETH Zurich, 8093 Zurich,
Switzerland}
\affiliation{$^2$Materials Science Division, Argonne National Laboratory, 
Lemont, IL 60439, USA}

\date{\today}

\begin{abstract}

Pinning and thermal creep determine the response of numerous systems
containing superstructures, e.g., vortices in type II superconductors, domain
walls in ferroics, or dislocations in metals. The combination of drive and
thermal fluctuations lead to the superstructure's depinning and its velocity
$v$ determines the electric, magnetic, or mechanical response. It is commonly
believed that pinning and creep collapse above the critical drive $F_c$,
entailing a sharp rise in the velocity $v$. We challenge this
perception by studying the effects of thermal fluctuations within the
framework of strong vortex pinning in type-II superconductors. In fact, we
show that pinning and thermal creep persist far beyond the critical force. The
resulting force-velocity characteristic largely maintains its zero-temperature
shape and thermal creep manifests itself by a downward renormalisation of the
critical drive. Such characteristics is in agreement with Coulomb's law of dry
friction and has been often observed in experiments.
\end{abstract}

\maketitle 

\section{Introduction}

The phenomenological behavior of numerous technological materials is
determined by topological defects, well known examples being vortices in
superconductors\cite{Blatter1994,Brandt1995}, dislocations in metals
\cite{Kassner2015,Zhou2015}, or domain walls in ferroic materials
\cite{Kleemann2007,Gorchon2014}.  Driving these topological objects via
suitable forces induces motion, with dramatic consequences for the material's
properties, e.g., loss of dissipation-free current transport in
superconductors, appearance of plastic flow in metals, or loss of magnetic
coercitivity in a ferromagnet.  Material imperfections come to rescue by
pinning these topological defects, vortices, dislocations, or domain walls, at
least up to a critical drive $F_c$ where pinning is finally overcome. Commmon
perception then tells that depinning, helped by thermal fluctuations, is a
dramatic effect that induces a steep onset of the superstructure's motion and
a rapid collapse of rigidity. In this paper, we demonstrate that such common
expectation is not generally applicable: assuming a strong pinning scenario
\cite{Labusch1969,LarkinOvchinnikov1979} applied to vortices in type II
superconductors with a small density of defects \cite{Blatter2004}, we
demonstrate that pinning and thermal creep persist far beyond the critical
drive, leading to a linear excess-current characteristic that is 
shifted by the action of thermal fluctuations.

\begin{figure} \centering \includegraphics[width=8.6cm]{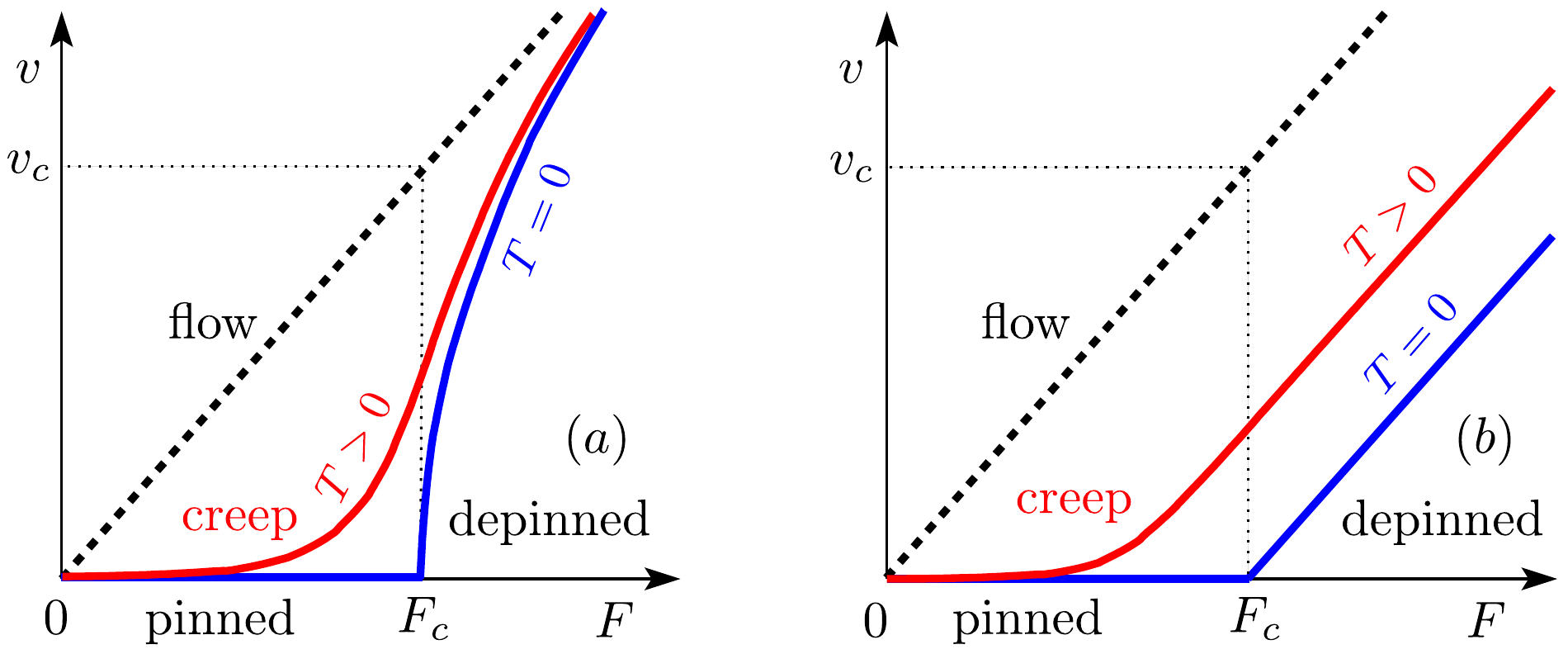}
\caption{\label{Fig:comparison} Velocity--force characteristics of a bulk
superconductor: (a) common perception with pinning force collapsing above
$F_c$, (b) calculated excess-force characteristic in accord with Coulomb's
law; $v_c = F_c/\eta$ denotes the velocity for free dissipative motion at
$F_c$.  Thermal creep appears mainly below $F_c$ in (a), while the persistence
of pinning beyond $F_c$ in (b) allows creep to manifest beyond $F_c$.}
\end{figure}

Given the ubiquitousness of the phenomenon, studies of the onset of motion of
pinned objects encompass a wide spectrum.  A simple but instructive setup is
given by a particle sliding down a tilted washboard potential. This model
describes the depinning of the superconducting phase and incipient voltage in
a current-driven Josephson junction \cite{AslamazovLarkin1969,Ambegaokar1969,
Tinkham} and has been used to describe the motion of flux bundles in a pinning
potential \cite{SchmidHauger1973,Tilley1990}: at depinning, the particle
dissipatively starts moving down the tilted washboard potential and the
velocity rises steeply, $v\propto (F-F_c)^{1/2}$, as pinning collapses beyond
$F_c$. Effects of thermal creep then are essentially limited to drives $F <
F_c$ below critical, see Fig.\ \ref{Fig:comparison}(a).  Dynamical
characteristics with steep velocity-onset as illustrated in Fig.\
\ref{Fig:comparison}(a) have become a common perception in drawing the shape
of a velocity--force characteristic, e.g., of superconducting material
\cite{Brandt1995,NattermannScheidl2000,Chauve2000,Giamarchi2006}.  In a
similar vein, effects of thermal creep are expected to manifest at drives $F <
F_c$.

However, this view contrasts with (classic) experimental data on bulk
superconducting material \cite{Strnad1964,Xiao2002} and recent theoretical
analysis \cite{Thomann2012} where the non-linear dynamical response assumes
the shape of an excess-force characteristic, see Fig.\
\ref{Fig:comparison}(b). This different shape is in agreement with Coulomb's
law of dry friction, telling that the static- and dynamical pinning forces are
equal and hence the pinning force persists even at drives beyond critical, $F
> F_c$.  Using strong pinning theory, we show that thermal effects produce a
downward shift of the critical- (or depinning-) force-density, while
preserving the shape of the excess-force characteristic, see Figs.\
\ref{Fig:comparison}(b) and \ref{Fig:characteristic}, confirming the presence
of pinning and its thermal reduction at large drives $F > F_c$.

\section{Strong pinning theory}

We consider a vortex lattice with density $a_0^{-2} = B/\Phi_0$, $\Phi_0 =
hc/2e$ the flux unit, induced by a field $B$ directed along the $z$-axis.  A
current density $j$ along $y$ drives these flux lines via the Lorentz-force
density $F_{\rm \scriptscriptstyle L} = j B/c$ along $x$.  Their free
dissipative motion $v = F_{\rm \scriptscriptstyle L}/\eta$, $\eta$ the
viscosity \cite{Bardeen1965}, is hugely modified by pinning due to material
defects, see Fig.\ \ref{Fig:comparison}.  Here, we use strong pinning theory
\cite{Labusch1969,LarkinOvchinnikov1979} in combination with Kramer's rate
theory \cite{Kramers1940} to determine the mean pinning-force density $\langle
F_\mathrm{pin}(v,T) \rangle$ opposing the vortex motion and study its
dependence on the creep velocity $v$ and temperature $T$. Using the result for
$\langle F_\mathrm{pin}(v,T) \rangle$ in the vortex dynamical equation
\begin{equation}\label{eq:veom}
   \eta v = F_{\rm \scriptscriptstyle L}(j) - \langle F_\mathrm{pin}(v,T)\rangle,
\end{equation}
we find the material's velocity--current ($v$--$j$) characteristic at finite
temperatures $T$ and evaluate the current-dependent barriers governing vortex
creep.

\begin{figure}
\centering \includegraphics[scale=1]{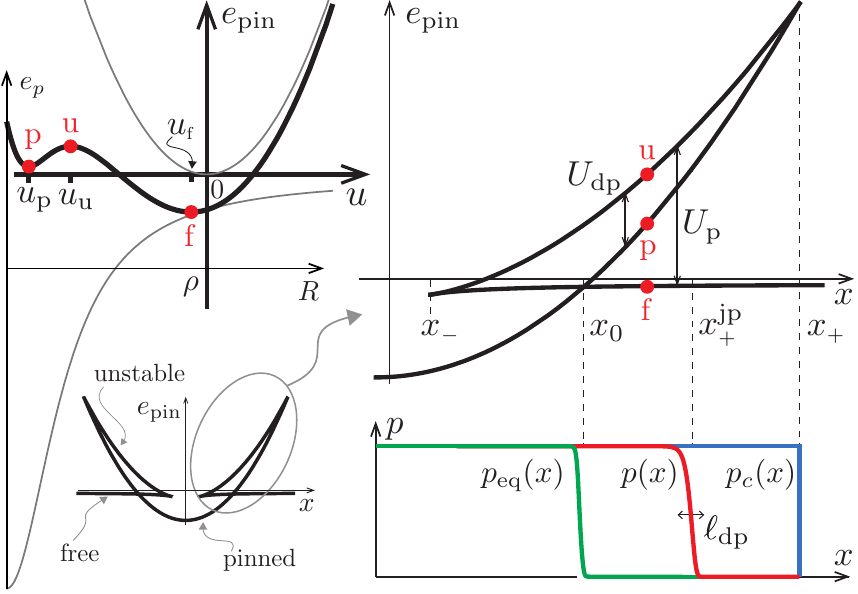}
\caption{\label{Fig:strong_pin} Top left: energy landscape $e_\mathrm{pin}
(u;\rho)$ versus $u$ (thick lines) as well as pinning potential $e_p(R)$
versus $R$ and elastic energy $\bar{C} u^2/2$ (thin lines). Shown is a
situation for $x_{\scriptscriptstyle -} < \rho < x_{\scriptscriptstyle{+}}$
with two local minima, pinned and free, with large and small distortions
$u_\mathrm{p} <0$ and $u_\mathrm{f} <0$. The unstable solution at
$u_\mathrm{u}$ defines the barrier separating the two minima. The bottom
sketch shows the pinning landscape $e_\mathrm{pin}(x)$ for a vortex driven
along the $x$-axis with free, pinned, and unstable branches at different
asymptotic positions $x$. Top right: expanded view of the region
$x_{\scriptscriptstyle -} < x < x_{\scriptscriptstyle{+}}$ with barriers
$U_\mathrm{dp}(x)$ and $U_\mathrm{p}(x)$ for depinning (vanishing $\propto
(x_{\scriptscriptstyle{+}} - x)^{3/2}$ at $x_{\scriptscriptstyle +}$) and
pinning (vanishing at $x_{\scriptscriptstyle -}$); the maximal barrier $U_0$
is attained at the branch crossing point $x_0$. Bottom right: branch
occupation $p(x;v,t)$ at $T=0$ and $v=0$ ($\to p_c$), at finite $T$ and $v$
($\to p$), and at equilibrium ($\to p_\mathrm{eq}$).}
\end{figure}

For a small pin density $n_p$ and defects that pin no more than one vortex,
the pinning problem can be reduced
\cite{Labusch1969,LarkinOvchinnikov1979,Blatter2004,Willa2016} to an effective
single-pin--single-vortex setup.  The latter involves a defect with a
potential $e_p({\bf R})\delta(z)$ of depth $e_p$ and extension $\xi$, the
coherence length, that we place at the origin. The vortex features an
effective elasticity $\bar{C} \sim \sqrt{\varepsilon_0 \varepsilon_l}/a_0$,
where $\varepsilon_l = \varepsilon_0 \ln (a_0/\xi)$ and $\varepsilon_0 =
(\Phi_0/4\pi\lambda)^2$ denote the vortex line elasticity and line energy,
respectively, and $\lambda$ is the screening length. With $\bar{C}$ related to
the local static elastic Green's function of the vortex lattice, we account
for the elastic forces of neighboring vortices
\cite{Labusch1969,LarkinOvchinnikov1979,Blatter2004,Willa2016}.  Assuming
defects of intermediate strength with $e_p/\xi < \varepsilon_0$ guarantees the
applicability of elasticity theory \cite{Thomann2017}.  Given an asymptotic
position $\boldsymbol{\rho}$ at large values of $|z|$, the vortex is locally
distorted by the presence of the defect, what results in a deformation ${\bf
u}$ within the plane $z = 0$. For a radially symmetric potential $e_p(R)$, the
problem further reduces to a scalar one involving only the radial asymptotic
distance $\rho$ of the vortex from the pin and the vortex displacement $u$
pointing towards the pin, hence $u < 0$.  The radial position $\rho + u$ of
the vortex tip can be found by minimizing the sum of pinning- and elastic
energies, see Fig.\ \ref{Fig:strong_pin},
\begin{align}\label{eq:e_pin}
   e_\mathrm{pin}(u;\rho) = e_p(\rho+u) + \bar{C} u^2/2
\end{align}
at fixed asymptotic position $\rho$, $\partial_u e_\mathrm{pin}(u;\rho) = 0$,
and we obtain the self-consistency equation for $u(\rho)$,
\begin{align}\label{eq:Cu_f}
   \bar{C} u(\rho) = f_p [\rho+u(\rho)], 
\end{align}
with $f_p(R) = - \partial_R e_p(R)$ the defect's force profile.  On the other
hand, the total derivative
\begin{align}\label{eq:f_pin}
   -\frac{d e_\mathrm{pin}[u(\rho);\rho]}{d\rho} 
   = f_p[\rho + u(\rho)] \equiv f_\mathrm{pin}(\rho)
\end{align}
provides us with the effective pinning force.

In the weak pinning situation, where the elasticity dominates, the nonlinear
self-consistency equation \eqref{eq:Cu_f} has a unique solution and pinning is
collective, involving many competing defects.  Strong pinning appears when the
Labusch parameter $\kappa \equiv \max_R[\partial_R f_p(R)]/\bar{C}$ is pushed
beyond unity, $\kappa > 1$. The total energy $e_\mathrm{pin}(\rho) \equiv
e_\mathrm{pin}[u(\rho);\rho]$ then exhibits multiple minima associated with
pinned [$u_\mathrm{p}(\rho)$] and free [$u_\mathrm{f}(\rho)$] vortex
configurations at the same asymptotic position $\rho$, see Fig.\
\ref{Fig:strong_pin}. Here, we consider strong pins with $\kappa > 1$ in the
presence of a small defect density with $n_p < (a_0 \xi^2 \kappa)^{-1}$,
implying less then one active pin per volume $a_0^3$---these conditions
delineate the three-dimensional strong-pinning regime in the $n_p$-$f_p$
diagram of Ref.\ [\onlinecite{Blatter2004}].

A current density $j$ along $y$ pushes the vortices along $x$ and we can
reduce the problem to a one-dimensional geometry. The pinning-force
density $\langle F_\mathrm{pin}(v,T) \rangle$ depends on the occupation
probability $p(x;v,T)$ of the pinned branch (the force along $y$ averages to
0),
\begin{align}\label{eq:F_pin_0}
   \langle F_\mathrm{pin}\rangle = -n_p \frac{2 t_{\scriptscriptstyle \perp}}{a_0} \! 
   \int \!\frac{dx}{a_0} \bigl[ p f^\mathrm{p}_\mathrm{pin} (x) 
   + (1 \! - \! p) f^\mathrm{f}_\mathrm{pin} (x) \bigr],
\end{align}
where $f^{\mathrm{f},\mathrm{p}}_\mathrm{pin}(x) \equiv
f_p[x+u_{\mathrm{f},\mathrm{p}}(x)]$ are the effective pinning forces
generated by the free and pinned branches and the integral is limited to the
interval $[-a_0/2,a_0/2]$ due to the periodicity of the lattice. At small
density $n_p$, different defects do not interact and thus $\langle
F_\mathrm{pin} \rangle \propto n_p$; furthermore, the average over $y$ can be
included with a factor $2t_{\scriptscriptstyle \perp}/a_0$, where
$t_{\scriptscriptstyle \perp}$ denotes the distance along $y$ over which
vortices get trapped \cite{Thomann2017}. At $T = 0$ and in the pinned state
with $v=0$, the maximally asymmetric occupation $p_c =
\chi(-x_{\scriptscriptstyle{-}}, x_{\scriptscriptstyle{+}})$ determines the
critical force density $F_c = \langle F_\mathrm{pin}(0,0) \rangle$, where
$\chi(a,b)$ denotes the characteristic function on the interval $[a,b]$ and
$\pm x_{\scriptscriptstyle \pm}$ are the boundaries of the pinned and free
branches; the condition $e_p/\xi < \varepsilon_0$ implies that $x_+ < a_0$ and
the periodicity of the vortex lattice does not interfere with the pinning
process.  Evaluating the integral in \eqref{eq:F_pin_0} with the help of
\eqref{eq:f_pin}, we obtain the critical force density
\begin{align}\label{eq:F_c}
   F_c = (2 x_{\scriptscriptstyle{-}}/a_0)\, n_p\, [\Delta e_\mathrm{p} + \Delta
   e_\mathrm{dp}]/a_0
\end{align}
with the jumps in energy $\Delta e_\mathrm{p} = [e^\mathrm{f}_\mathrm{pin}-
e^\mathrm{p}_\mathrm{pin}]_{x=-x_{\scriptscriptstyle{-}}}$ upon pinning at
$-x_{\scriptscriptstyle{-}}$, $\Delta e_\mathrm{dp} =
[e^\mathrm{p}_\mathrm{pin} - e^\mathrm{f}_\mathrm{pin}]_{x
=x_{\scriptscriptstyle{+}}}$ at depinning, and $e^{\mathrm{f},
\mathrm{p}}_\mathrm{pin} (x) \equiv e_\mathrm{pin}[u_{\mathrm{f},
\mathrm{p}}(x);x]$.  For a radially symmetric pinning potential, vortices
approaching the defect jump into the pin at a distance $\rho =
x_{\scriptscriptstyle{-}}$ and hence the transverse trapping length is given
by $t_{\scriptscriptstyle \perp} = x_{\scriptscriptstyle{-}}$.

\section{Thermal fluctuations}

At finite temperatures $T > 0$, inspired by the work on charge-density-wave
pinning \cite{BrazovskiLarkin1999,BrazovskyNattermann2004}, see also Ref.\
\cite{Fisher1985}, we can find the branch occupation probability $p(x;v,T)$
within the bistable regions $x_{\scriptscriptstyle{-}} < |x| <
x_{\scriptscriptstyle{+}}$ from the rate equation (note that $p(|x| >
x_{\scriptscriptstyle{+}}) = 0$ and $p(|x| < x_{\scriptscriptstyle{-}}) = 1$)
\begin{align}\label{eq:rate}
   \partial_t p = v \partial_x p = -p\,\omega_\mathrm{p}\, e^{-U_\mathrm{dp}/T}
   + (1-p)\,\omega_\mathrm{f}\, e^{-U_\mathrm{p}/T}.
\end{align}
The barriers $U_\mathrm{p,dp}$ are determined by the third solution
$u_\mathrm{u}(x)$ of Eq.\ \eqref{eq:Cu_f} which is unstable, see Fig.\
\ref{Fig:strong_pin}, $U_\mathrm{p}(x) = e^{\mathrm{u}}_\mathrm{pin}(x) -
e^{\mathrm{p}}_\mathrm{pin}(x)$ and $U_\mathrm{dp}(x) =
e^{\mathrm{u}}_\mathrm{pin}(x) - e^{\mathrm{f}}_\mathrm{pin}(x)$. The attempt
frequencies $\omega_{\mathrm{p},\mathrm{f}}(x)$ relate to the curvatures of
the total energy $e_\mathrm{pin}(u;x)$ at the extremal points and account for
the dissipative vortex dynamics \cite{Kramers1940}.

\subsection{Large drives}

For {\it large drives} $F_{\rm \scriptscriptstyle L} \sim F_c$, the occupation
probability $p(x;v,T)$ maintains its steps, albeit smoothed due to thermal
fluctuations and shifted to new positions $x^\mathrm{jp}_{\scriptscriptstyle
\pm}(v,T)$ where vortices jump between free and pinned branches, see Fig.\
\ref{Fig:strong_pin}. In determining the depinning point
$x^\mathrm{jp}_{\scriptscriptstyle{+}} (v,T) < x_{\scriptscriptstyle{+}}$, we
focus on the first term in Eq.\ \eqref{eq:rate}. We define the local
relaxation length $\ell_\mathrm{dp}(x) \equiv [v/\omega_\mathrm{p}(x)]\,
e^{U_\mathrm{dp}(x)/T}$ and take another derivative of Eq.\ \eqref{eq:rate} to
obtain the curvature $\partial_x^2 \, p \approx (p/\ell_\mathrm{dp}^2)(1 +
\ell'_\mathrm{dp})$. We define the jump position through the inflection point,
i.e., $\partial_x^2 \, p\, (x^\mathrm{jp}_{\scriptscriptstyle{+}}) = 0$, and
arrive at the condition $\ell_\mathrm{dp} (x^\mathrm{jp}_{\scriptscriptstyle
{+}}) \approx T/|U'_\mathrm{dp} (x^\mathrm{jp}_{\scriptscriptstyle +})|$ for
$x^\mathrm{jp}_{\scriptscriptstyle{+}}$, with $f'(x)$ the derivative of $f(x)$
and we have ignored the $x$-dependence of $\omega_\mathrm{p}$.  The criterion
for the pinning point $-x^\mathrm{jp}_{\scriptscriptstyle {-}}$ is derived
from an analogous consideration with $\ell_\mathrm{p} = (v/\omega_\mathrm{f})
e^{U_\mathrm{p}/T}$ replacing $\ell_\mathrm{dp}$. Defining the thermal
velocity scale \cite{v_th_precise}
\begin{align}\label{eq:v_th}
   v_\mathrm{th} = {\omega_\mathrm{p} T}/{|U'_\mathrm{dp} (x^\mathrm{jp}_{\scriptscriptstyle +})|}
   \sim \kappa^s{\omega_\mathrm{f} T}/{|U'_\mathrm{p} (x^\mathrm{jp}_{\scriptscriptstyle{-}})|}
\end{align}
with $s = (n+3)/(n+2)$ depending on the decay $e_p(x) \propto x^{-n}$, we can
cast these criteria into the simple form
\begin{align}\label{eq:U_v}
   U_\mathrm{dp}(x^\mathrm{jp}_{\scriptscriptstyle +}) \approx U_\mathrm{p}
   (x^\mathrm{jp}_{\scriptscriptstyle{-}})
   \approx T \ln (v_\mathrm{th}/v).
\end{align}
These results are valid for barriers $U_\mathrm{dp,p} \gg T$, i.e., for
velocities $v$ small compared to $v_\mathrm{th}$. As $v$ approaches
$v_\mathrm{th}$ at large drives $F_{\rm \scriptscriptstyle L} > F_c$,
$x^\mathrm{jp}_{\scriptscriptstyle \pm} \to x_{\scriptscriptstyle \pm}$, the
barriers $U_\mathrm{dp,p}$ vanish, and the characteristic approaches the $T=0$
result.

\subsection{Small drives}

At {\it small drives}, the jump locations $x^\mathrm{jp}_{\scriptscriptstyle
\pm}$ approach the branch crossing point $x_0$ where the barrier $U_0 =
U_\mathrm{dp}(x_0)$ $ = U_\mathrm{p}(x_0)$ is maximal, see Fig.\
\ref{Fig:strong_pin}. Pinning and depinning transitions become equally
important and the probability $p(x;v,T)$ differs perturbatively from the
equilibrium occupation $p_\mathrm{eq}(x) = [1+\ell_\mathrm{p}(x)/
\ell_\mathrm{dp}(x)]^{-1}$. The rate equation \eqref{eq:rate} can be rewritten
in the form $\partial_x p = (p_\mathrm{eq} - p)/\ell_\mathrm{eq}$, with the
equilibrium relaxation length $\ell_\mathrm{eq}(x)$ given by
$\ell^{-1}_\mathrm{eq} = \ell^{-1}_\mathrm{p} + \ell^{-1}_\mathrm{dp}$; its
solution takes the form of a right-shifted equilibrium occupation, $p(x)
\approx p_\mathrm{eq} [x-\ell_\mathrm{eq}(x)]$.

\section{Response Characteristic}

The branch occupation probabilities $p(x;v,T)$ determine the effective
pinning-force density via Eq.\ \eqref{eq:F_pin_0}.  In addition, for small
creep velocities, we have $t_{\scriptscriptstyle \perp}(v,T) =
x^\mathrm{jp}_{\scriptscriptstyle{-}}$, with a saturation at $x_0$ as $v \to
0$.  Finally, the velocity $v$ is found from a self-consistent solution of the
vortex equation of motion \eqref{eq:veom}.  Below, we carry out this program
and determine the superconductor's $v$--$j$ characteristic that is shown Fig.\
\ref{Fig:characteristic}.

\subsection{Large drives}

\begin{figure}[t]
\centering
\includegraphics[scale=1]{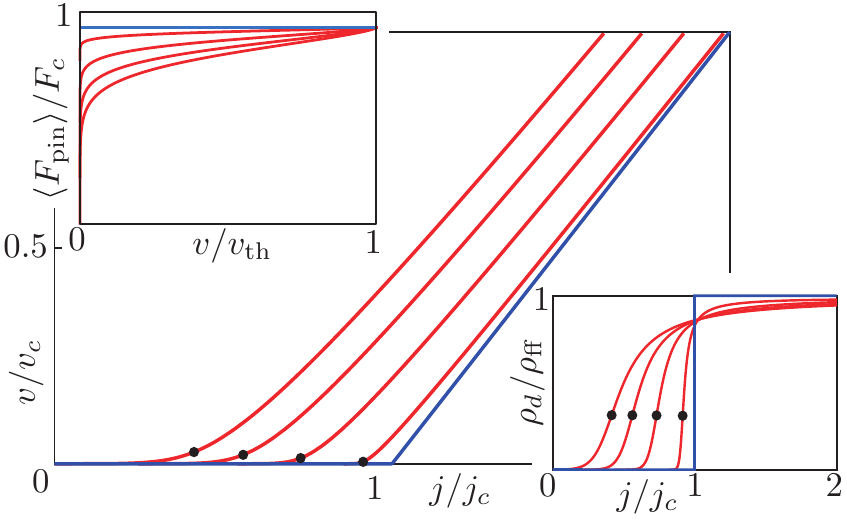}
\caption{$v$--$j$ characteristic at temperatures $T/e_p = (0, 0.1,$ $0.5,
\,1.0,\,1.5)\times 10^{-2}$ and for a small defect density $n_p a_0 \xi^2 =
10^{-4}$; we have chosen a Labusch parameter $\kappa = 5$ implying $g(\kappa)
\approx 2.8$ and $a(\kappa) \approx 0.17$ for a Lorentzian potential $e_p(R)$.
Thermal fluctuations lead to a downward shift of $j_c$ to $j_\mathrm{dp}(T)$,
with the latter (solid points) defined through the inflection point in
$\rho_d(j)$, see lower-right inset.  The weak logarithmic dependence on $v$ of
$\rho_d$ at currents $j > j_c$ closely preserves the shape of the excess-current
characteristic also at finite $T$, with creep manifesting itself well above
$j_c$.  The top-left inset shows the pinning-force density $\langle
F_\mathrm{pin}(v,T>0)\rangle/F_c$, reduced due to thermal creep for velocities $v <
v_\mathrm{th}$. Linear TAFF response is not visible on this scale.}
\label{Fig:characteristic}
\end{figure}

We first consider large drives $F_{\rm \scriptscriptstyle L} \sim F_c$. Given the
small width $\ell_\mathrm{dp} \sim (T/e_p)\, x_{\scriptscriptstyle +}$ of the
jump in the occupation probability $p(x; v,T)$, see Fig.\
\ref{Fig:strong_pin}, we can use the approximation $p(x;v,T) \approx
\chi(-x^\mathrm{jp}_{\scriptscriptstyle{-}}, x_{\scriptscriptstyle
+}^\mathrm{jp})$ in Eq.\ \eqref{eq:F_pin_0} and obtain the pinning-force
density
\begin{align}\label{eq:F_pin}
   \langle F_\mathrm{pin}(v,T)\rangle = (2 x^\mathrm{jp}_{\scriptscriptstyle{-}}/a_0)
   \,n_p \, [\Delta e^\mathrm{jp}_\mathrm{p} 
   + \Delta e^\mathrm{jp}_\mathrm{dp}]/a_0
\end{align}
with the reduced jumps $\Delta e^\mathrm{jp}_\mathrm{p}$ and $\Delta
e^\mathrm{jp}_\mathrm{dp}$ evaluated at the positions $x =
-x^\mathrm{jp}_{\scriptscriptstyle{-}}$ and $x =
x^\mathrm{jp}_{\scriptscriptstyle +}$, cf.\ Eq.\ \eqref{eq:F_c}.  Expanding
Eq.\ \eqref{eq:F_pin} for small deviations $\delta x_{\scriptscriptstyle \pm}
= \pm (x_{\scriptscriptstyle \pm} - x^\mathrm{jp}_{\scriptscriptstyle \pm})
>0$ and normalizing, we obtain the force-density ratio
\begin{align}\label{eq:force_expansion}
   \frac{\langle F_\mathrm{pin}(v,T) \rangle}{F_c} 
   = 1+ \frac{\delta x_{\scriptscriptstyle{-}}} {x_{\scriptscriptstyle{-}}}
   - \frac{\Delta e'_\mathrm{p} \delta x_{\scriptscriptstyle{-}}
   +\Delta e'_\mathrm{dp} \delta x_{\scriptscriptstyle +}} 
   {\Delta e_\mathrm{p}+\Delta e_\mathrm{dp}},
\end{align}
where $\Delta e_\mathrm{p}'$ and $\Delta e_\mathrm{dp}'$ denote derivatives of
$\Delta e_\mathrm{p}$ and $\Delta e_\mathrm{dp}$ at
$-x_{\scriptscriptstyle{-}}$ and $x_{\scriptscriptstyle +}$, respectively.
The first (positive, since $x^\mathrm{jp}_{\scriptscriptstyle{-}} >
x_{\scriptscriptstyle{-}}$) correction is due to the change in the trapping
distance $t_{\scriptscriptstyle \perp}$, while the second term represents the
decrease in the pinning-force density due to the reduced asymmetry in the
branch occupation.  Assuming a smooth pinning potential $e_p(x)$ of depth
$e_p$ and large $\kappa$, one finds \cite{Buchacek2018} that
$U_\mathrm{dp}(x^\mathrm{jp}_{\scriptscriptstyle +}) \sim e_p (\delta
x_{\scriptscriptstyle +}/\kappa \xi)^{3/2}$ and
$U_\mathrm{p}(x^\mathrm{jp}_{\scriptscriptstyle{-}}) \sim e_p \kappa^s (\delta
x_{\scriptscriptstyle{-}}/\kappa \xi)^{3/2}$.  Using these results with
$U_\mathrm{p} \approx U_\mathrm{dp}$ in Eq.\ \eqref{eq:force_expansion}, we
find that
\begin{align}\label{eq:Fpin}
   \langle F_\mathrm{pin}(v,T)\rangle / F_c \approx 1- g(\kappa)(U_\mathrm{dp}/e_p)^{2/3}, 
\end{align}
with $g(\kappa) = \tilde{g}(\kappa)[\kappa/(\kappa - 1)]^{4/3}$ collecting all
prefactors of $\delta x_{\scriptscriptstyle \pm}$ and $\tilde{g}(\kappa)$
depending on the shape of $e_p(x)$, $\tilde{g}(\kappa)$ of order 2 for a
Lorentzian shaped potential \cite{Buchacek2018} $e_p(R) = e_p/(1+R^2/2\xi^2)$.
Combining this result with Eq.\ \eqref{eq:U_v}, the equation of motion
\eqref{eq:veom} assumes the simple form
\begin{align}\label{eq:charac}
   v/v_c = j/j_c - 1 + g(\kappa) (T/e_p)^{2/3}
   \, [\ln(v_\mathrm{th}/v)]^{2/3}
\end{align}
that involves the critical current density $j_c =c F_c/B$ and two velocity
scales, the flux-flow velocity at $F_c$, $v_c = F_c/\eta \propto n_p$, and the
thermal velocity $v_\mathrm{th}$, see Eq.\ \eqref{eq:v_th}.  The $v$--$j$
characteristic is easily obtained by plotting $j(v)$, see Fig.\
\ref{Fig:characteristic}.  At $T=0$, we recover the linear excess-current
characteristic \cite{Thomann2012} with $v = v_c\, (j/j_c-1)$ for current
densities $j>j_c$.  The effect of thermal fluctuations is conveniently
analyzed via the differential resistivity scaled with the free flux-flow
resistivity $\rho_\mathrm{ff} \propto v_c/j_c$,
\begin{align}\label{eq:djv}
   \frac{\rho_d}{\rho_\mathrm{ff}} \equiv \frac{d (v/v_c)}{d (j/j_c)} = \biggl[1
   + \frac{2\tau^{2/3}}{3} \frac{v_c/v}{[\ln(v_\mathrm{th}/v)]^{1/3}}\biggr]^{-1},
\end{align}
where we have defined the rescaled temperature $\tau = g^{3/2}(\kappa) \,
T/e_p.$ As illustrated in Fig.\ \ref{Fig:characteristic}, $\rho_d$ expressed
through $j$ assumes a step-like form that is shifted to lower currents as $T$
increases.  We define the depinning current-density $j_\mathrm{dp}(T)$ through
the inflection point $\partial_j^{\,2} \rho_d = 0$; assuming a large ratio
$\alpha = (v_\mathrm{th}/v_c) \, \tau^{-2/3}$, we find that
\begin{align}\label{eq:jdp}
    j_\mathrm{dp}(T) \approx j_c \bigl[1- \tau^{2/3} \{ \ln[3 \alpha\, (\ln
    3\alpha)^{1/3}]\}^{2/3}\bigr]
\end{align}
and $\rho_d(j_\mathrm{dp}) \approx \rho_\mathrm{ff}/3$. The velocity ratio
$v_\mathrm{th}/v_c = (T/e_p) a(\kappa)/n_p a_0 \xi^2$ involves another factor
$a(\kappa) = \tilde{a}(\kappa)$ $\kappa^{-1/(n+2)} [\kappa/(\kappa - 1)]^{3/2}$
that depends on $e_p(x)$, with $\tilde{a}(\kappa)$ of order 0.1 for a
Lorentzian potential \cite{Buchacek2018}.

The rounding of the $v$--$j$ characteristic near $j_\mathrm{dp}$ is
conveniently described by a creep barrier $U(j) \equiv U_\mathrm{dp}
[v(j),T]$; approximating the equation of motion \eqref{eq:veom} $\langle
F_\mathrm{pin} \rangle/F_c \approx j/j_c$ for small velocities $v$ and using
\eqref{eq:Fpin}, we find a creep-type motion $v \approx v_\mathrm{th} \, e^{-U(j)/T}$
with a barrier
\begin{align}\label{eq:Uj<}
   U(j \lesssim j_c) \approx e_p[(1-j/j_c)/g(\kappa)]^{3/2}.
\end{align}
The most important feature of the $v$--$j$ characteristic in Fig.\
\ref{Fig:characteristic} is the persistence of creep far beyond $j_c$. This is
very different from a characteristic describing a rapid collapse of the
pinning-force density beyond $j_c$ with a steep rise in velocity $v$ at $j_c$
and thermal creep prevailing below $j_c$, see Fig.\ \ref{Fig:comparison}(a).
For strongly-pinned vortices, the pinning-force density $\langle
F_\mathrm{pin}(v)\rangle$ persists for drives beyond $j_c$; such behavior
coincides with Coulomb's law of dry friction that is at the origin of the
excess-current characteristic \cite{Thomann2012}.  Since $\langle
F_\mathrm{pin}(v)\rangle$ survives $j_c$, depinned vortices still profit from
thermal activation and creep manifests itself beyond $j_c$.  Furthermore,
changes in the pinning-force density $\langle F_\mathrm{pin} (v)\rangle$ are
logarithmic in $v$ and hence small, giving rise to a flat resistivity
$\rho_d(j)$ above $j_c$.  As a result, the $v$--$j$ characteristic is
renormalized downwards but keeps an excess-current form at finite
temperatures, see Fig.\ \ref{Fig:characteristic}. Finally, the characteristic
joins the $T=0$ excess-current characteristic at $j_\mathrm{th} = j_c(1+
v_\mathrm{th}/v_c)$ where $x^\mathrm{jp}_{\scriptscriptstyle \pm} \to
x_{\scriptscriptstyle \pm}$ and the pinning-force density $\langle
F_\mathrm{pin}(v_\mathrm{th})\rangle = F_c$, see Fig.\
\ref{Fig:characteristic}, with the velocity ratio $v_\mathrm{th}/v_c \propto
T/n_p$ attaining large values for small defect densities $n_p$.

\subsection{Small drives}

We find the pinning-force density $\langle F_\mathrm{pin}(v,T) \rangle$ at
small velocities $v\lesssim v_{\rm \scriptscriptstyle TAFF} = v_\mathrm{th}
e^{-U_0/T}$ (i.e., small drives $F_{\rm\scriptscriptstyle L} \ll F_c$) by
inserting the shifted equilibrium distribution $p_\mathrm{eq}[x -
\ell_\mathrm{eq}(x)]$ into Eq.\ \eqref{eq:F_pin_0}.  Expanding  in small
$\ell_\mathrm{eq} \propto v$ and making use of the anti-symmetry
$f_\mathrm{pin}^{\mathrm{p},\mathrm{f}}(x) =
-f_\mathrm{pin}^{\mathrm{p},\mathrm{f}}(-x)$, we obtain
\begin{align}\label{eq:F_pin_small_v}
   \langle F_\mathrm{pin}\rangle \approx - n_p \frac{2 x_0}{a_0}\!\!
   \int \! \frac{d x}{a_0}\ell_\mathrm{eq}(x)\,p_\mathrm{eq}'(x)\,
   \Delta f_\mathrm{pin}(x)
\end{align}
with $\Delta f_\mathrm{pin}(x) = f_\mathrm{pin}^\mathrm{f}(x) -
f_\mathrm{pin}^\mathrm{p}(x)$.  A simple estimate is obtained by replacing
$p_\mathrm{eq}'(x)$ with a sum of $\delta$-functions at $\pm x_0$, see Fig.\
\ref{Fig:strong_pin}; accounting for the precise shapes of $p_\mathrm{eq}(x)$
and $\ell_\mathrm{eq}(x)$ contributes a $\kappa$-dependent prefactor.  Using
$\ell_\mathrm{eq}(x_0) = v\, (\omega_\mathrm{p}+\omega_\mathrm{f})^{-1}
e^{U_0/T}$ and the scalings $\omega_{\mathrm{p},\mathrm{f}}\sim
(e_p/\xi^2)/\eta a_0^3$, $x_0 \sim \xi$, and $\Delta f_\mathrm{pin}(x_0)\sim
e_p/\xi$, we arrive at the result
\begin{align}\label{eq:F_pin_TAFF}
   \langle F_\mathrm{pin}(v,T)\rangle = \eta v \, h(\kappa) (n_p a_0 \xi^2)
   \,e^{U_0/T},
\end{align}
with the barrier $U_0 = e_p \tilde{u}(\kappa)[(\kappa-1)/\kappa]^2$ and all
$\kappa$-depen\-dence absorbed in $h(\kappa) = \tilde{h}(\kappa)
\kappa^{(n+2)/(4n+4)} [\kappa/(\kappa-1)]^{1/2}$; for a Lorentzian potential,
we find $\tilde{h}(\kappa)$ of order 20 and $\tilde{u}(\kappa)$ of order 0.3
(for $\kappa = 5$, we have $h(\kappa) \approx 36$ and $U_0/e_p \approx 0.18$).
At low temperatures, Eq.\ \eqref{eq:F_pin_TAFF} implies a TAFF characteristic
with an exponentially suppressed slope as compared to free flux-flow,
\begin{align}\label{eq:j_v_TAFF}
   \frac{v}{v_c} = \frac{j}{j_c} \, \frac{e^{-U_0/T}}{h(\kappa)\, n_p a_0\xi^2}.
\end{align}
The crossover to the non-linear characteristic is realized at the velocity
$v_{\rm \scriptscriptstyle TAFF}$ corresponding to the driving current $j_{\rm
\scriptscriptstyle TAFF} \approx a(\kappa)\, h(\kappa)(T/e_p) \,
j_c$.

\section{Conclusion}

In conlusion, we have shown that, contrary to usual expectation, thermal creep
persists far beyond the critical depinning current density $j_c$ when pins are
dilute and strong. This unexpected result is in accord with the excess-current
characteristic following from Coulomb's law.  Such a characteristic and its
temperature dependence is easily set apart from the steep characteristic
associated with the collapse of the pinning force beyond $j_c$, possibly due
to the avalanche-type depinning conjectured for weakly-pinned random elastic
manifolds \cite{NarayanFisher1992,Chauve2000}.  These insights provoke further
work directed at understanding the crossover between the dry-friction type
characteristic typical for diluted strong pins and the collapse-type
characteristic usually associated with dense weak pins.  Finally, strong
pinning theory provides a quantitative result for the linear TAFF response at
small currents, see Eq.\ \eqref{eq:j_v_TAFF}.  The latter has been
experimentally observed and quantitatively analyzed, e.g., in high temperature
superconductors \cite{Iye1987,Palstra1988,Tinkham1988}; conversely, the
downward shift and rounding of the excess-current characteristic predicted by
strong pinning theory awaits more detailed experimental and numerical
investigations.

\bigskip

\acknowledgements

We thank A.E.\ Koshelev for discussions and acknowledge financial support
from the Swiss National Science Foundation through the Division II, the
National Centre of Competence in Research `MaNEP--Materials with Novel
Electronic Properties' and an Early Postdoc.Mobility Fellowship (R.W.).
The work at Argonne was supported by the US Department of Energy, Office of
Science, Materials Sciences and Engineering Division.


\begin{thebibliography}{99}
\bibitem{v_th_precise} Close to $x_{\scriptscriptstyle +}$ the velocity scale
$v_\mathrm{th}$ assumes a constant value $v_\mathrm{th} =
(\kappa_{\scriptscriptstyle +}/2\pi)(T/\eta a_0^3\xi)$ with
$\kappa_{\scriptscriptstyle +} = \xi f_p''[x_{\scriptscriptstyle
+}+u_\mathrm{p}(x_{\scriptscriptstyle +})]/2\bar{C} \sim \kappa$, where the
last relation is valid for very strong pinning.
\end{thebibliography}
\end{document}